\documentclass[amsmath,amssymb,prl,twocolumn,showpacs]{revtex4} 
\usepackage{graphicx} 
\usepackage{dcolumn} 
\usepackage{bm} 
\begin{document} 

\title{Strong cooperativity and inhibitory effects in DNA multi-looping 
processes} 

\author{Artur Garc\'\i a-Saez$^{1}$} 
\author{J. Miguel Rubi$^{2}$} 
\affiliation{$^{1}$ICFO-Institut de Ciencies Fotoniques, Mediterranean 
Technology Park, 08860 Castelldefels (Barcelona), Spain.\\ 
$^{2}$Departament de Fisica Fonamental, Facultat de Fisica, Universitat de Barcelona, Diagonal 647, 08028 Barcelona, Spain.} 


\begin{abstract} 
We show the existence of a high interrelation between the different loops that 
may appear in a DNA segment. Conformational changes in a 
chain segment caused by the formation of a particular loop may either 
promote or prevent the appearance of another. The underlying loop selection mechanism 
is analyzed by means of a Hamiltonian model from which the looping free energy and 
the corresponding repression level can be computed. 
We show significant differences between the probability of single and multiple loop formation.
The consequences that these collective effects might have on gene regulation processes are outlined.  
\end{abstract} 

\maketitle 

Loop formation in DNA complexes has been identified as a 
fundamental mechanism in gene 
regulation processes \cite{switch,data1,data2,widom,vilar2}. Operators for 
DNA-protein interaction modify their relative positions through the formation of loops and thereby
may operate even if they are not physically close together. This mechanical 
process is governed by the physical properties of the DNA and the 
concentration of proteins and has a deep impact on 
gene synthesis processes. The fixation of operators combined with  
protein concentrations is responsible for control processes inside 
the cell. 

Elastic models have been proposed for the study of the physical 
properties of the DNA chain and the emerging phenomena 
like cyclisation and looping \cite{metzler,purohit}, and are the basis for
large scale simulations of protein complexes \cite{villa}.
Within this approach, the elasticity of the bonds between the nucleotid bases
determine the physical properties
of DNA through its degrees of freedom. 
An important step toward the understanding of looping phenomena within a 
physical context was given in \cite{vilarPRL}, where
the effect of protein concentration was related to multiprotein bonding 
positions. An induced phase transition to the loop phase is controlled 
by the protein concentration. 
Following this physical analysis, a model of loop formation has been 
proposed using ideas from statistical mechanics which provides a clear 
picture of the connection between the protein concentrations, 
the free energy involved in loop formation \cite{vilar2} and protein binding, as well as 
the structure of the DNA. The transition between the loop formation 
phase was reported for the case of a single loop and multiple proteins. 

In this Letter, we show that the loop selection process is the result of a 
strong competition between the different types of loops that can be formed in 
the same DNA fragment. These loops may appear in DNA segments with several
binding configurations, but also in single loop configurations with 
the possibility of different spatial dispositions of the looped segment \cite{gelles}. 
Loop formation entails changes in the structure of the DNA chain
which allow distal operators to come into range of a binding protein (see Fig.1). 
However, in a scenario where 
multiple loops may appear, these conformational changes can hamper or even promote additional loop creation 
once the appearance of a loop has modified the conditions necessary 
for the formation of additional loops. The possibility of formation 
of multiple loops becomes manifest through an effective 
interaction between loops that may for instance affect their size \cite{markoPRL}.

\begin{figure} 
\includegraphics[width=0.95\linewidth]{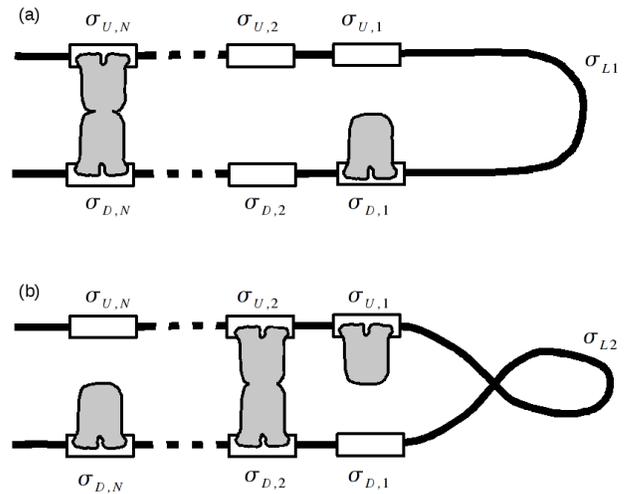} 
\caption{(a) Loops are formed in a DNA segment by protein binding. The DNA 
chain is looped by the interaction of 
binded protein monomers fixed at corresponding distant sites. We track this binding using 
the binary variables $\sigma_{U,i}, \sigma_{D,i}$. 
$\sigma_{L_k}$ marks the formation of the loop $k$. 
(b) Different loops may appear due to alternate protein configurations 
or DNA spatial disposition.} 
\end{figure} 

We focus on the formation of competing types of loops in 
a segment of DNA assuming that only one loop may be present at the same time in 
the segment. The
conditions necessary for the formation of a loop are either geometrical, where 
the required operators have been 
set in positions that are incompatible with additional loop formation, 
or energetic, where 
the energy to form another loop is not strong enough to undo an existing loop. 
In general, the most energetically favorable loops will be dominant; however, other loops may also emerge due to the interaction of the proteins 
binded to the chain during loop formation.
As a result, a conformational interaction is induced between potential loops.

Loop formation due to the binding of multiple proteins can be put in a 
statistical mechanics language by means of a Hamiltonian model which 
reflects the successive steps intervening in the process \cite{vilarPRL}. 
In a DNA segment with $2N$ binding positions with $M$ different loops, 
the corresponding Hamiltonian can be written as 
\begin{eqnarray} 
H & = & 
\sum_{k=1}^{M}\left[\sigma_{L_k}\left(c_{k}+\sum_{i=1}^{N}e_{k}\sigma_{U,i}
\sigma_{D,i}\right)\right. 
\nonumber\\ 
 &  & 
\left.+\sum_{i=1}^{N}\left(g_{U,i}\sigma_{U,i}+g_{D,i}\sigma_{D,i}\right)\right]
\;.\label{Ham2loop} 
\end{eqnarray} 
Here the set of binary variables $\sigma_{L,k}$ (=0,1) accounts for the 
formation of 
a type $L_k$ loop, and the variables 
$\sigma_{U,i}$ and $\sigma_{D,i}$ indicate the binding of a protein monomer at 
the corresponding position (see Fig.1). 
The contributions to the free energy for the formation of a loop are 
introduced through the coefficients $c_{k}$  
(which are independent of the chain length), while the coefficients $e_{k}$ 
on the other hand, multiply the number of dimers $\sigma_{U,i}\sigma_{D,i}$ contributing to 
loop formation which can be a function
of the chain length.
Different types of loops may carry different values of $c_k$ and $e_{k}$. 
The coefficients $g_{U,i}$ and $g_{D,i}$ are associated to 
the contributions of binding a monomer to the chain. Throughout this work we set 
$g_{U,i}=g_{D,i}=g=g_o-\frac{1}{\beta}\ln n$, where the protein concentration $n$ is introduced in
the Hamiltonian, and the binding contribution $g$ is site independent.

\emph{Two-loop interaction-} 
We focus our analysis on the case $M = 2$ which shows the basic features of 
loop interactions. An additional study of 
cases with $M>2$ has revealed the absence of 
important differences in the loop selection mechanism. Changes in the chain due 
to the formation of a loop $L_1$ modify the conditions under which another 
potential configuration of a looped phase $L_2$ may emerge. This 
situation can be found in short chains where the deformation of the 
DNA after 
the formation of a loop alters the distance and possible contact between distal 
monomers. 
We then envisage a scenario where loops of different free 
energies of formation compete. Once one 
of the loops is formed, there is no room for others. 
This restriction can be mathematically expressed as 
\begin{equation} 
\sum_{k=1}^M \sigma_{L_k} \leq 1.
\label{eq:rest} 
\end{equation} 

\begin{figure} 
\includegraphics[width=0.97\linewidth]{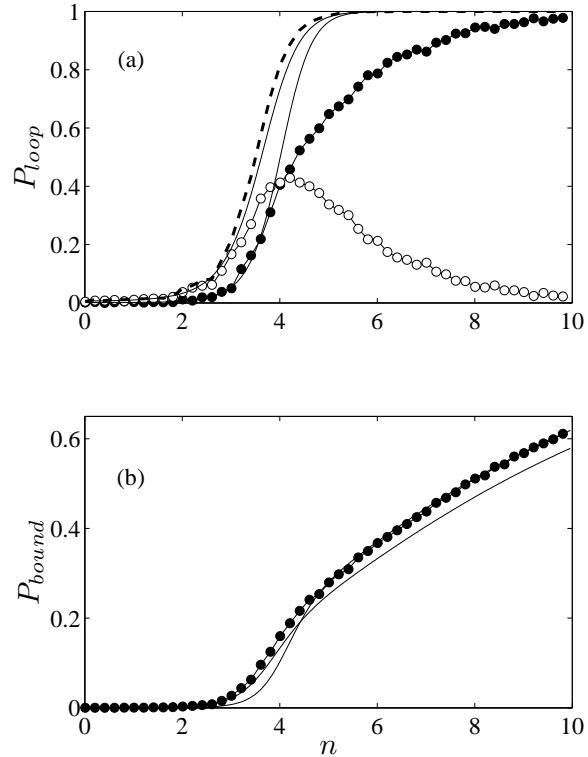} 
\caption{\label{fig2}(a) Loop formation probability for two coexisting 
loops with $N=30$, $c_1 = 4,2$ kcal/mol, $e_1=-8$ kcal/mol, $c_2 = 3$ kcal/mol and $e_2=-7,9$ kcal/mol. 
The solid lines show the probability for the case in which only one loop can be 
formed while the marks are the result for the case of two loops. The dashed
line shows the total probability of the looped phase. 
(b) Binding probability of a protein monomer in the 
DNA, in the same conditions. These proteins contribute to the formation of the two loops. The 
solid lines show the binding probability for the one-loop case.} 
\end{figure} 

By using Montecarlo methods, the probabilities $P_{loop}(L_k)=\langle \sigma_{L_k} 
\rangle$ and $P_{bound}=\langle \sigma_{U/D,i} \rangle$ 
can be computed from the resulting equilibrium states. Analytical results can be
obtained for the single-loop scenario, and are used as a reference for
the multiple-loop results shown here.
The protein concentration $n$ is the order parameter which describes the 
transition between the looped/unlooped phases \cite{vilarPRL} of the different $L_k$. 
To analyze this transition, we deal with $M+2$ body interactions corresponding to 
the interaction of two operator sites to form one loop and the restriction 
imposed over the $M$ loops. Adding the restriction 
Eq.\ref{eq:rest}, we study the values $P_{loop}(L_1)$ and 
$P_{loop}(L_2)$ for a chain with two possible loops with different free energy contributions. 

We start by studying the formation of two loops $L_1$ and $L_2$ in a 
chain with $N=30$, $c_1 = 4,2$ kcal/mol, $e_1=-8$ kcal/mol, $c_2 = 3$ kcal/mol and $e_2=-7,9$ kcal/mol. 
We set $g_0 = -7.2$ kcal/mol and $\beta^{-1}=0.6$ kcal/mol in all our computations.
In Fig.\ref{fig2}, we show the results for the probabilities of loop 
formation (top) and the probability of binding a monomer (bottom). 
The solid lines represent the expected values of the 
probability of single-loop formation in the absence of interaction, taken from \cite{vilarPRL}. 
The formation of multiple loops can be analyzed similarly through that of a single-loop with an effective interaction. The marks 
show the corresponding results of the Montecarlo simulation. 
  
Under these conditions, one of the two loops appears only in a small range of the 
protein concentration $n$. Thus, the activity of the cell processes associated 
with the formation of this loop is restricted to this range of concentrations, 
making induced loop interaction a mechanism for gene control inside the cell. 
This behavior is produced by the two different contributions 
to the free energy of the loop formation, 
given through the term $c_k$, independent of the chain size, and the term $e_k$ which depends on the number of protein dimers present in the 
chain. This contribution depends on the protein concentration $n$ inside the cell, 
becoming greater for higher values of $n$. Thus, a loop 
may become dominant at low $n$ due to a dominant constant contribution 
$c_k$. By increasing the protein concentration, the free energy contribution of 
term $e_k$ becomes dominant due to the formation and binding of more dimers 
contributing to loop formation. This mechanism changes the corresponding loop probability of the 
different types of loops (see Fig.\ref{fig2}(a)). The binding probability of the 
monomers gets contributions from the two forming loops, thus becoming the basic 
mechanism behind loop interaction. In Fig.\ref{fig2}(b), $P_{bound}$ is equal to that of the 
dominant loop for high $n$, while for $n\sim 3-4$, it receives contributions from 
the two loops. 

\begin{figure}[ht] 
\includegraphics[width=0.97\linewidth]{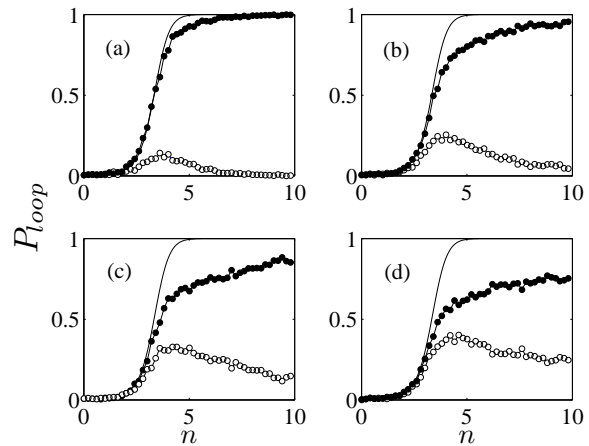} 
\caption{\label{fig3} Loop formation probability $P_{loop}$ for two 
coexisting loops with 
$N=30$, $c_1 = c_2 = 3$ kcal/mol, $e_1=-8$ kcal/mol and 
$e_2=-7.9$  kcal/mol (a), $-7.95$  kcal/mol (b), $-7.97$ kcal/mol 
(c), $-7.98$ kcal/mol (d). 
The solid lines show the results for the one-loop case, 
while the marks show the 
results of the Montecarlo simulation.} 
\end{figure}

We have extended the interaction study to a range of values of $e_1$ and $e_2$ 
for which the coexisting loop picture goes from an equiprobable 
disposition of both loops (fixing $c_1 = c_2$) to a situation where one of the loops dominates. For 
$e_2 = e_1 +\Delta e$, 
with increasing $\Delta e$, we identify the transition region where the 
probability $P_{loop}(L_2)$ is zero for high protein concentrations. 
The results are shown in Fig.\ref{fig3}. This transition 
depends nontrivially on the respective values of $e_1$ and $\Delta e$ and 
the protein concentration $n$ and shows a progressive inhibition of $L_2$ 
formation for increasing $\Delta e$. The $L_2$ formation is restricted 
to a progressively  narrow range of values 
of $n$, making this mechanism a way to activate some cell processes for 
very particular 
protein concentrations. As explained above, this fact is a consequence of the 
dimer formation that contributes to the formation of 
the loop. 
The dimer concentration increases with $n$ which can be interpreted as 
the contribution to the free energy of the dimer formation. 

\begin{figure}[ht] 
\includegraphics[width=0.90\linewidth]{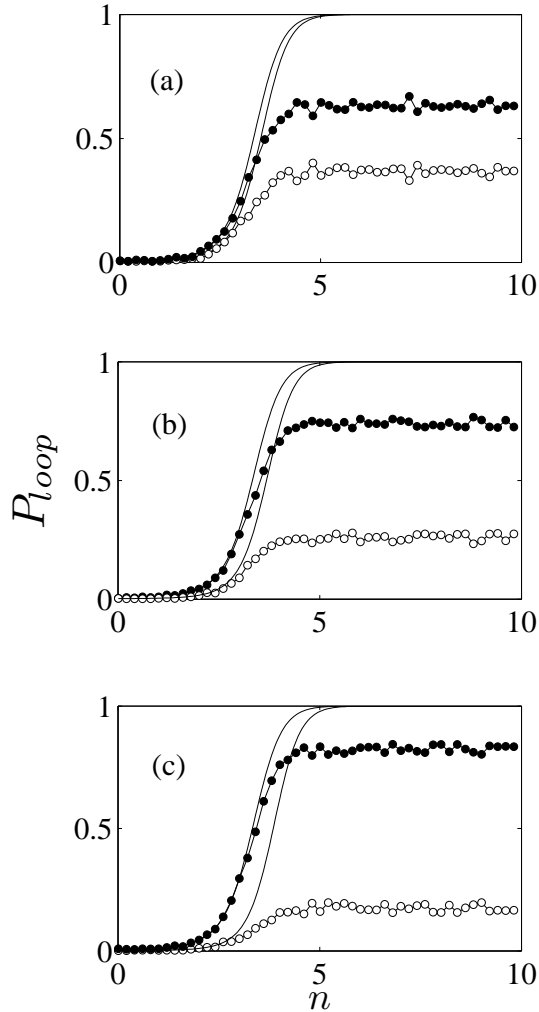} 
\caption{\label{fig4} Loop formation probability $P_{loop}$ for two 
coexisting loops with $e_1= e_2 =-8$  kcal/mol, $c_1=3$ kcal/mol and
$c_2 = 3.3$  kcal/mol (a), $3.6$  kcal/mol (b), $3.9$  kcal/mol(c). 
The solid lines show the single-loop values (in isolated conditions), 
while the marks show the result of the Montecarlo simulation.} 
\end{figure}

We will now analyze the case of two loops with $e_1=e_2$ 
and $c_1=c_2+\Delta c$. 
The transition in this case is driven by a constant contribution to the 
Hamiltonian independent of $n$. 
In Fig.\ref{fig4}, we show $P_{loop}$ for different values of $\Delta c$. 
As expected, for high values of 
$n$ there is no variation in $P_{loop}$ after the transition, 
resulting in the same relative probabilities for the two loops at different protein 
concentrations. This behavior is of a
completely different nature from that shown in Fig.\ref{fig3} and can be 
interpreted as the contribution 
of the different structures of DNA to the free energy. This situation may appear
in loops with different potential physical dispositions, with different values of $c_k$, 
but formed in equivalent conditions of dimer bonding.

\emph{Repression level-} We have computed the probabilities of two types of loops
$L_1$ and $L_2$ in a single DNA chain. In physiological conditions where 
the looped phase is associated with the 
repression of a gene (\emph{i.e.} the \emph{lac} operon in 
\emph{E.coli} \cite{lac}), an effective looped phase
probability $P_{L_{eff}}$ can be computed. Under some conditions this probability is higher 
than the respective probabilities in the single-loop case (see Fig.2(a)): $P_{loop}(L_1)$
and $P_{loop}(L_2)$. 
The probability of a sate is determined by its standard free energy $H_k$
through $P_k \propto e^{-H_k/RT}$, normalized by the probabilities of all the possible 
configurations. Thus the \emph{effective} free energy of the looped phase $H_{L_{eff}}$
satisfies
\begin{equation} 
H_{L_{eff}} < H_{L_1}, H_{L_2}
\end{equation} 
where $H_{L_1}$ and $H_{L_2}$ are the corresponding free energies of the looped
phases $L_1$ and $L_2$.

The contributions to the free energy of the DNA molecule can be identified 
with the repression levels \cite{vilar-leibler,rpnas}. The free 
energy of the DNA molecule can be computed 
from the different contributions of binding and loop formation. 
Hence we can connect this physical interpretation with the experimental 
measurements of the repression levels. Taking into consideration the \emph{lac} repression mechanism, 
the repression level $R_{loop}$ with a single looped phase is given by \cite{rpnas}
\begin{equation}
R_{loop}= 1 + e^{-g/RT}\left([N] + e^{-H_{L_1}/RT}\right).
\end{equation} 
The repression level in loop interaction conditions $\tilde{R}_{loop}$, considering the effective
loop free energy contribution and Eq.(3), satisfies
\begin{equation} 
\tilde{R}_{loop} \gtrsim R_{loop}. 
\end{equation} 
The repression of transcription induced by the loop formation, in situations where 
multiple loop formation can appear, is affected by the corresponding conditions of protein 
concentrations and loop properties. Repression levels in the single-loop 
scenario have been reported in \cite{data1,data2}.

\emph{Conclusions-} 
We have shown the presence of strong correlations between the different 
loops that can be formed in a given DNA segment. Geometrical 
changes in the chain, caused by the formation of a loop, can alter  
the conditions under which another loop may come up, thereby implying 
modifications of the loop formation probability and consequently of
their statistical properties.
These correlations can give rise to cooperative effects for which loops may 
appear under otherwise forbidden conditions and to inhibitory effects hampering 
the loop formation under apparently favorable conditions. The loop interrelation effect 
can be quantified through an effective free energy which can be computed from 
a Hamiltonian that incorporates all the energies coming into play in the process. 
These collective effects can be adapted to a 
wide combination of physical conditions inside the cell, where small 
changes of the protein concentrations can dramatically alter the 
cellular processes controlling the repression level. 
The implications that loop collective effects may have in gene regulation 
processes can then be studied from measurable quantities 
establishing a clear connection 
between the repression level and the possible loop configurations in a fragment of DNA.

 

\end{document}